\documentclass{PoS}

\usepackage{amsmath}
\usepackage{amssymb}
\usepackage{graphicx}
\usepackage[T1]{fontenc} 
\usepackage{slashed}
\usepackage{xspace}
\usepackage{multirow}
\usepackage{cite}

\newcommand\MG{{\tt MadGraph}\xspace}
\newcommand\SARAH{{\tt SARAH}\xspace}

\newcommand\FeynArts{{\tt FeynArts}\xspace}
\newcommand\FormCalc{{\tt FormCalc}\xspace}
\newcommand\CalcHep{{\tt CalcHep}\xspace}

\newcommand\MicrOmegas{{\tt MicrOmegas}\xspace}
\newcommand\WHIZARD{{\tt WHIZARD}\xspace}

\newcommand\SPheno{{\tt SPheno}\xspace}
\newcommand\Vevacious{{\tt Vevacious}\xspace}

\newcommand\HB{{\tt HiggsBounds}\xspace}
\newcommand\HS{{\tt HiggsSignals}\xspace}
\newcommand\SSP{{\tt SSP}\xspace}

\newcommand\FlexibleSUSY{{\tt FlexibleSUSY}\xspace}
\newcommand\LanHEP{{\tt LanHEP}\xspace}
\newcommand\FeynRules{{\tt FeynRules}\xspace}
\newcommand\FlavorKit{{\tt FlavorKit}\xspace}

\newcommand{\T}{{\bf \,\hat T}}
\newcommand{\PH}{{\bf \,\hat \chi}}
\newcommand{\Oc}{{\bf \,\hat O}}
\newcommand{\Ta}{{\bf \,\hat T_1}}
\newcommand{\Tb}{{\bf \,\hat T_2}}
\newcommand{\Si}{{\bf \,\hat S}}
\newcommand{\Hd}{{\bf \,\hat H_d}}
\newcommand{\Hu}{{\bf \,\hat H_u}}
\newcommand{\Rd}{{\bf \,\hat R_d}}
\newcommand{\Ru}{{\bf \,\hat R_u}}

\title{Beyond-MSSM Higgs sectors}
\ShortTitle{Beyond-MSSM Higgs sectors }

        \author{\speaker{Florian Staub}\\
       Bethe Center for Theoretical Physics \& Physikalisches Institut der 
Universit\"at Bonn, \\
 53115 Bonn, Germany\\
       E-mail: \email{fnstaub@th.physik.uni-bonn.de}}

\abstract{
This is a compact overview of Higgs sectors in extensions of the MSSM. The focus is on 
the summary of the main 
features of models with additional singlets and triplets as well as of models with Dirac 
gauginos. In addition, also important aspects of models with an extended gauge sector 
are shown. Finally, I comment on available tools which can be used for an adequate study 
of non-minimal SUSY models. }

\FullConference{Prospects for Charged Higgs Discovery at Colliders - CHARGED 2014,\\
		16-18 September 2014\\
		Uppsala University, Sweden}

\begin{document}

\section{Introduction}
Supersymmetry (SUSY) has been the top candidate for beyond standard model (BSM) physics since many years. 
While in the past the focus has been on the minimal supersymmetric standard model (MSSM), the null results 
from LHC for SUSY searches \cite{Craig:2013cxa} as well as the rather large Higgs mass \cite{Chatrchyan:2012ufa,Aad:2012tfa}
have triggered more interest in SUSY models
beyond the MSSM. The reason is that BMSSM model could not only address these problems but might 
answer also other questions which still remain open in the MSSM. An incomplete list of motivations to 
go beyond the MSSM is the following:
\begin{itemize}
 \item {\bf Naturalness}: the need to push the Higgs mass to the observed level by large loop corrections gets significantly
 softened if additional $F$- or $D$-term contributions to the tree-level mass are present
 \cite{Ellwanger:2009dp,Ellwanger:2006rm,Ma:2011ea,Zhang:2008jm,Hirsch:2011hg}.
 \item {\bf Missing SUSY signals}: the unsuccessful searches  for SUSY have put impressive limits on the SUSY masses in the simplest manifestation 
 of SUSY. 
 However, in the context of compressed spectra or $R$-parity violation these limits become much weaker \cite{Dreiner:2012gx,Bhattacherjee:2013gr,Kim:2014eva}.
 \item {\bf Neutrino masses}: neutrinos would still be massless in the MSSM. To incorporate neutrino masses, either $R$-parity has to be 
 violated to allow for a mixing of the neutrinos with SUSY states or additional particles are needed which contribute to the neutrino masses \cite{Borzumati:2009hu,Rossi:2002zb,Hirsch:2008dy,Esteves:2009vg,Esteves:2010ff,Malinsky:2005bi,%
Abada:2012cq,BhupalDev:2012ru,Abada:2014kba}.
 \item {\bf $\mu$-problem}: the $\mu$ parameter in the superpotential from the MSSM must be of $O(EWSB)$ because of phenomenological reasons. However, 
 since it is not protected by any symmetry its natural size would be  $O(GUT)$. To relax this tensions, $\mu$ could be generated dynamically
 as a consequence of SUSY breaking like in singlet extensions \cite{Kim:1983dt,Ellwanger:2009dp}.
 \item {\bf Strong CP-problem}: also the question about the strong CP problem remains open in the MSSM. To solve it, one can introduce a Peccei-Quinn
 symmetry \cite{Peccei:1977hh}. The minimal, self-consistent SUSY model doing that needs three additional superfields whose scalar components 
  can mix with the MSSM Higgs states \cite{Dreiner:2014eda}.
 \item {\bf UV-completion}: there are 
 many SUSY scenarios motivated by GUT or string models where additional gauge groups are broken close to the TeV scale. 
 These models predict usually plenty of additional states   close to the SUSY scale beside $Z'$ and $W'$.
\end{itemize}
Most extensions of the MSSM have 
in common that they come together with an extended Higgs sector. I'll give therefore an overview about the most popular extensions 
of the MSSM Higgs sector in the next section before I comment in sec.~\ref{sec:tools} on tools which can be used to study these 
and many other models.

\section{Overview about non-minimal Higgs sectors}
Extending the Higgs sector of the MSSM can have several consequences:
(i) additional contributions to the Higgs mass can be present;
(ii) the MSSM doublets mix with other states what will change the character of the 'SM-like' Higgs boson;
(iii) as consequence of this mixing the couplings of {\it the} Higgs can be modified compared to SM expectations;
(iv) additional light scalars with a reduced couplings to SM particle can be present;
(v) additional charged and also double charged bosons can appear. 

What happens and how important an effect is, depends on the concrete model.  Therefore, I'm going to discuss briefly 
the most important MSSM extensions in the following. I categorize the extension into two groups:
(i) models with the SM gauge sector, (ii) models with an extended gauge sector, and start with the first one. 

\subsection{Models with SM gauge sector}
I start with models which don't extend the gauge sector of the MSSM and consider in this case 
in particular singlet and triplet 
extensions as well as models with Dirac gauginos. Of course, 
there are many models which  I'll have to skip, e.g. the DiracNMSSM \cite{Lu:2013cta,Kaminska:2014wia}, 
models with a PQ-symmetry \cite{Dreiner:2014eda}, models with bilinear $R$-parity
 violation \cite{Hirsch:2000ef,Hundi:2013lca}, sister Higgs models \cite{Alves:2012fx}, 
 models with a gauged $R$-symmetry \cite{Chamseddine:1995gb} and many more. 
I'll always assume that the superpotential is decomposed as $W = W_Y + W_X$, where $W_Y$ contains 
the Yukawa and $W_X$ the Higgs part for a given model. In the MSSM $W_X$ corresponds to
\begin{equation}
W_{MSSM} =  \mu \Hd \Hu \, . 
\end{equation}

\subsubsection{Singlet extensions}
The simplest ansatz to go beyond the MSSM is to add a superfield which is a gauge singlet. The general superpotential 
for the Higgs sector with all renormalizable terms allowed by gauge invariance reads
\begin{equation}
W_S = t_S \Si + \mu_S \Si^2 + \kappa \Si^3 + \mu \Hd \Hu + \lambda \Si \Hd \Hu \, .
\end{equation}
Usually, one proposes a discrete symmetry to forbid some of the these terms. The most studied assumption 
is the next-to-minimal supersymmetric standard model (NMSSM) with a  $Z_3$ which forbids all dimension-full parameters: 
$t_S=\mu_S=\mu = 0$, see \cite{Ellwanger:2009dp,Ellwanger:2006rm} and references therein. 
Other possibilities are the near-to-minimal SSM (nMSSM) with a $Z^R_5$ ($\mu_S=\mu = \kappa = 0$)
\cite{Panagiotakopoulos:1999ah,Panagiotakopoulos:2000wp}
and the 
general NMSSM (GNMSSM) with a $Z^R_8$ ($t_S=0$) \cite{Lee:2010gv, Lee:2011dya,Ross:2011xv}. 
All realizations have in common that they predict additional $F$-term contributions
to the tree-level Higgs to evade the condition $m_h^{Tree} < m_Z$ known from the MSSM.
The tree-level mass in singlet extensions can be approximated as
\begin{equation}
 m_h^{2,Tree} = m_Z^2 \cos^2 2\beta + \frac{\lambda^2}{2} v^2 \sin^2 2\beta \, ,
\end{equation}
and in the limit of very small $\tan\beta < 3$ and large $\lambda > 0.5$ one finds that $m_h^{Tree}$ can be easily above 
100~GeV because of the second term. $\lambda$ is usually assumed to be below $0.65$ to have a theory 
which is perturbative up to the GUT scale. If this is given up, even larger $\lambda$ couplings 
are possible \cite{Hall:2011aa}. 
The enhanced tree-level mass relaxes significantly the necessity of large loop corrections via (s)tops and renders such models a more natural candidate 
for BSM physics. One can quantify the naturalness of a model  with respect to a set of independent parameters, $p$,
by considering a fine-tuning (FT) measure like \cite{Ellis:1986yg, Barbieri:1987fn}
\begin{equation} 
\label{eq:measure}
\Delta \equiv \max {\text{Abs}}\big[\Delta _{p}\big],\qquad \Delta _{p}\equiv \frac{\partial \ln
  v^{2}}{\partial \ln p} = \frac{p}{v^2}\frac{\partial v^2}{\partial p} \;.
\end{equation}

\begin{figure}[hbt]
\centering
\includegraphics[width=0.5\linewidth]{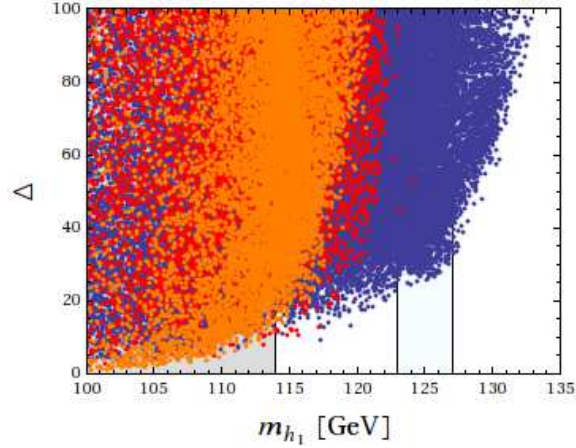} 
\caption{Fine-tuning $\Delta$ in the MSSM (orange) and GNMSSM (blue) in a fully constrained model. 
Plot taken from Ref.~\cite{Ross:2012nr}}
\label{fig:FT}
\end{figure}

It has been shown 
that the NMSSM improves significantly the FT compared to the MSSM
\cite{BasteroGil:2000bw,Dermisek:2005gg,Dermisek:2006py,Dermisek:2007yt,Ellwanger:2011mu}. 
Moreover, going to the GNMSSM reduces the FT even further
\cite{Ross:2011xv,Ross:2012nr,Kaminska:2013mya}, see also Fig.~\ref{fig:FT} for a comparison 
of the FT in a constraint version of the MSSM and GNMSSM as function of $m_h$. 

From a 
phenomenological point of view already the extension by just one singlet superfield can have profound consequences: new decay 
channels for the SM-like Higgs can appear (e.g. $h \to A A / H H \to 4 b / 4 \tau / 2 b 2 \tau$) \cite{Ellwanger:2001iw,Stal:2011cz,King:2014xwa}. 
Also couplings to SM particles can be altered significantly: there is 
for instance the possibility to change the effective  $h \gamma \gamma$ coupling 
either due to a mixing
with the singlet or by chargino loops enhanced by large $\lambda$ \cite{Ellwanger:2011aa,SchmidtHoberg:2012yy,SchmidtHoberg:2012ip,King:2012is}. 
If the singlinos and the singlet have masses of only a few GeV, 
this can help to hide SUSY at the LHC because it reduces the missing transversal energy ($\slashed{E}_T$) to a level below the one needed for 
many SUSY searches \cite{Ellwanger:2014hia}. On the other side new search strategies for charged Higgs fields are possible by considering the cascade
$t\to b H^- \to b W^- H/A \to b W^- \gamma \gamma$ \cite{Das:2014fha}.

\subsubsection{Triplet extensions}
In the case of triplet extensions, one can consider either a model with only one 
triplet which doesn't carry hypercharge ($\T$, $Y=0)$ or a model with 
two triplets with hypercharge ($\Ta, \Tb$, $Y=\pm 1$).
The different terms in the superpotential of the two models read \cite{Basak:2012bd,Kang:2013wm,Arina:2014xya,Bandyopadhyay:2014tha}
\begin{eqnarray}
W_{T1} &=& \mu_T \text{Tr}(\T^2) + \lambda_T \Hd \T \Hu + \mu \Hu \Hd \, , \\
W_{T2} &=& \mu_T \text{Tr}(\Ta \Tb) + \lambda_u \Hu \Ta \Hu  + \lambda_d \Hd \Tb \Hd + \mu \Hu \Hd \, .
\end{eqnarray}
In general, triplet extensions share many features with singlet extensions: there is a $F$-term enhancement 
to the Higgs mass, the Higgs branching ratios can be affected by the presence of the new particle(s) and new cascade decays
compared to the MSSM can arise. A feature compared to singlet extensions is the presence of additional charged Higgs bosons. For 
the model with two triplets even double-charged Higgs bosons appear. Finally, triplet extensions affect the $\rho$ 
parameter what constraints the parameter values in this kind models. If the triplets get
 non-vanishing vacuum expectation values (VEVs), 
$\rho$ is already shifted at tree-level, i.e. triplet VEVs must be very small. At one-loop one finds limits 
on combinations of $(\lambda_T, \mu_T)$ \cite{DiChiara:2008rg}. Unfortunately, both kinds of triplet extensions
spoil the nice feature of gauge coupling unification.

\subsubsection{Singlet/Triplet extensions}
One can also combine the two ideas and  consider a model with triplets and a singlet
at the same time \cite{Agashe:2011ia}: 
\begin{eqnarray}
W_{ST} &=& \lambda_T \Si \text{Tr}(\Ta \Tb)+ \lambda_S \Si \Hu \Hd + \kappa \Si^3  + \lambda_u \Hu \Ta \Hu + \lambda_d \Hd \Tb \Hd   \, .
\end{eqnarray}
The advantage of this setup is that the $\mu$-problem for the triplets gets solved, too. In addition, one 
can easier keep $\delta\rho$ under control. \\ 
It's a matter of taste if this is enough motivation 
to assume the presence of both extensions. 
However, there is also a kind of models which {\it predicts} the presence of singlets and triplets instead of adding them 
ad-hoc: SUSY models with Dirac gauginos, which I'm going to discuss now. 

\subsubsection{Models with Dirac gauginos}
In general, there are two possibilities to generate mass terms for gauginos $\lambda$:
\begin{equation}
 M_M \lambda \lambda \hspace{1cm} M_D \lambda \Psi
\end{equation}
$M_M$ is a Majorana mass term, while $M_D$ is a Dirac mass term due to the interaction with a superfield $\Psi$ in 
the adjoint representation \cite{Fox:2002bu,Nelson:2002ca,Davies:2011mp,Benakli:2008pg,Benakli:2011kz}. 
Dirac mass terms are theoretical well motivated because they are a consequence
of $N=2$ SUSY. In contrast to Majorana masses Dirac masses are also consistent with an $R$-symmetry. Thus, 
if one assumes an underlying $R$-symmetry which forbids Majorana masses, singlet, triplet and octet superfields are 
needed to generate masses for all gauginos. Not only the Majorana masses are forbidden by the $R$-symmetry, 
but also trilinear soft-terms  as well as bilinear terms in the superpotential. These constraints give this kind of 
models a new character compared to the extensions before because one adds not only new properties but also 
forbids feature of the MSSM. Therefore, models with Dirac gauginos can differ significantly from the MSSM:
(i) the cross sections of colored SUSY states can be suppressed by the Dirac character of the gluino 
\cite{Choi:2008ub,Kribs:2009zy,Choi:2010gc,Kribs:2012gx}; (ii) the constraints 
from flavor physics get relaxed \cite{Kribs:2007ac, Dudas:2013gga};
(iii) because of the supersoftness of the theory the RGEs especially for scalar soft 
masses change significantly and the mass pattern appearing in a constrained model are completely different to those in the CMSSM
\cite{Goodsell:2012fm, Benakli:2014cia,Busbridge:2014sha}. 

\paragraph*{Broken R-Symmetry in Higgs sector}
If one just adds the superfields $\Si$, $\T$ and $\Oc$ which are necessary to 
generate Dirac gaugino masses, $R$-symmetry in the Higgs 
sector has to be broken. This happens by assuming that a subset of the following, $R$-symmetry violating terms is present
\cite{Benakli:2012cy}
\begin{equation}
W_{\slashed{R}} =  (\lambda_S \Si + \mu) \Hu \Hd   +  \lambda_T \Hd \T \Hu + \kappa \Si^3 \, .
\end{equation} 
Even if these terms violate $R$-symmetry they neither introduce Majorana masses nor trilinear soft-terms. Therefore, the 
radiative corrections of (s)tops to the Higgs are largely suppressed compared to the MSSM with large stop mixing. Thus, it is either 
necessary to have very heavy stops in the multi TeV range or to enhance the Higgs mass already at tree-level via the 
additional $F$-term known
from the NMSSM by choosing large $\lambda$ and small $\tan\beta$. 
Another consequence of Dirac mass terms is the presence of new $D$-terms of the form $M_D \tilde{\Psi}^a \phi^* T^a \phi$ 
($T^a$ are the generators of the gauge groups). The corresponding $U(1)_Y$, $SU(2)$ terms 
give negative contributions to the tree-level Higgs mass. Thus, the bino and wino Dirac mass is usually assumed not to be too large.

\paragraph*{Unbroken R-Symmetry}

\begin{figure}[hbt]
\centering
\vspace{-2em}
\includegraphics[width=0.75\linewidth]{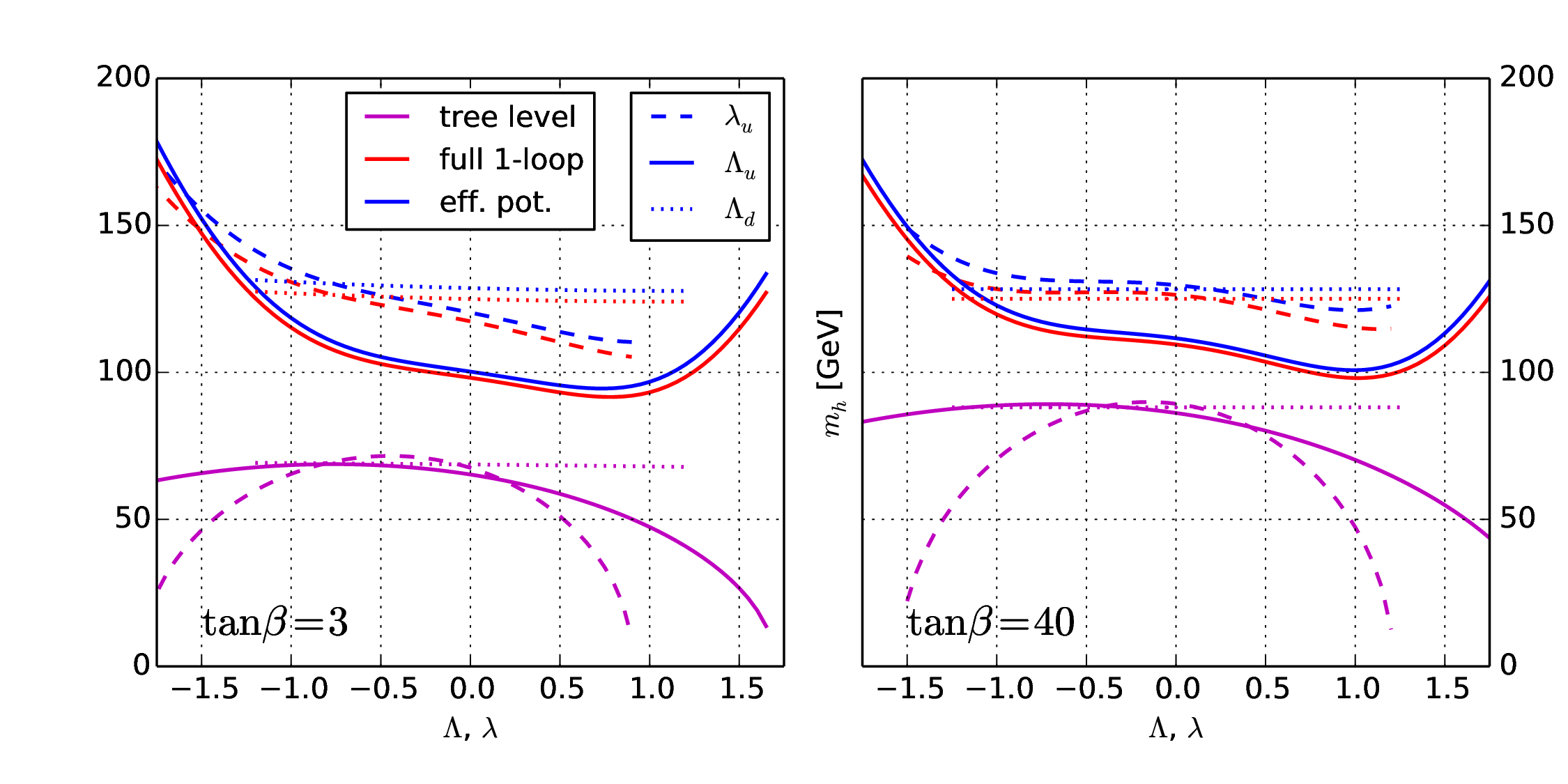} 
\caption{Lights Higgs mass in the MRSSM at tree-level and one-loop as 
function of the new $\lambda$/$\Lambda$ couplings in the superpotential. Plots
are an updated version of the ones of Ref.~\cite{Kotlarski:2014} and were kindly provided by Wojciech Kotlarski.
More results are given in Ref.~\cite{Diessner:2014ksa}.}
\label{fig:MRSSM}
\end{figure}

If $R$-symmetry is taken to be unbroken, the Higgs sector has to be extended by two doublets $\Ru$ and $\Rd$ 
which allow to write down $R$-symmetric $\mu$-terms \cite{Kribs:2007ac}. 
\begin{eqnarray}
W_R &=& (\mu_u + \lambda_u \Si) \Hu \Ru +  (\mu_d + \lambda_d \Si) \Hd \Rd  + \Lambda_d \Rd \T \Hd + \Lambda_u \Ru \T \Hu \, .
\end{eqnarray}
This is the minimal-$R$-symmetric SSM (MRSSM) and it has many additional differences compared to the MSSM. For instance, it predicts an 
asymmetric dark matter candidate because the neutralinos are also Dirac states. Since there is no $\lambda$-term to 
enhance the Higgs mass, the tree-level mass is usually lighter than in the MSSM \cite{Bertuzzo:2014bwa}
 \begin{eqnarray}
m_h^2 &&\simeq M_Z^2 \cos^2 2\beta  - v^2 \left(\frac{(g_1 M_D^B + \sqrt{2}\lambda \mu)^2}{(4(M_D^B)^2+m_s^2)}+  \frac{(g_2 M_D^W + \Lambda \mu)^2}{(4(M_D^W)^2+m_T^2)}\right)
 \end{eqnarray}
This effect together with the reduced (s)top corrections is not necessarily a big problem as one might think: loop corrections proportional 
to $\lambda_{d,u}$ or $\Lambda_{d,u}$ can be used to push the Higgs mass to 125~GeV as shown in Fig.~\ref{fig:MRSSM}.
Also in the charged Higgs sector this model is very interesting:
it predicts not only three charged Higgs particles but also two additional charged $R$-Higgs fields. The phenomenology of these new charged states 
is hardly explored at the moment. 

\subsection{Models with extended gauge sector}
Extending the SM gauge sector 
\begin{equation}
G_{SM}=SU(3)_c\times SU(2)_L \times U(1)_Y
\end{equation}
introduces not only additional gauge bosons but also scalars to break the new gauge group. 
The easiest extensions are those with a single $U(1)$. However, GUT theories like $SO(10)$ 
predict often also additional $SU(N)$ groups and in string models multiple $U(1)$'s can 
be present. These additional groups can be in principle be broken at any scale close to or significantly above the 
TeV range. I'm going to concentrate here on scenarios where this breaking happens at energies 
which can be probed in the near future directly at colliders. For higher breaking scales only indirect probes 
like from flavor observables are possible \cite{Esteves:2010si}.

\subsubsection{$U(1)$ extensions}
\begin{figure}[hbt]
\vspace{-1em}
\includegraphics[width=0.49\linewidth]{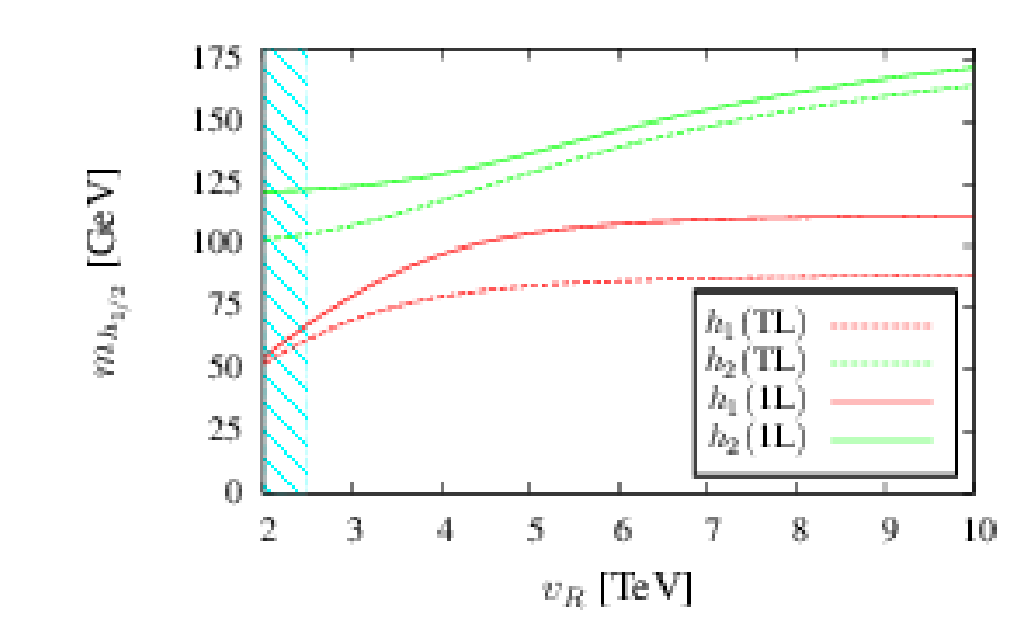} \hfill
\includegraphics[width=0.49\linewidth]{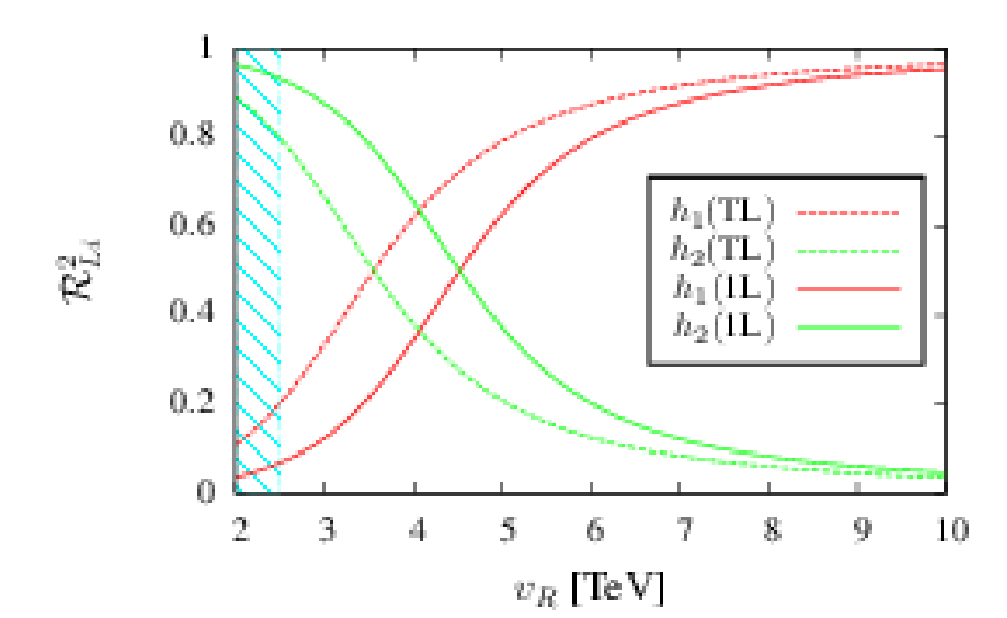} 
\caption{Masses (left) and doublet-fraction (right) of the two lightest
scalars in a $U(1)_R \times U(1)_{B-L}$ model at tree-level and one-loop. 
Plots taken from Ref.~\cite{Hirsch:2011hg}}
\label{fig:U1R}
\end{figure}

There are many different realizations for $G_{SM} \times U(1)_X$ with 
$X = \chi, R,  B-L, N, \eta, Y, S, I, \slashed{p},  \dots$, see for instance Refs.~\cite{Erler:2010uy}
and references therein. The concrete 
gauge group is often fixed by the underlying string or GUT theory one has in mind. 
The kind of $U(1)$ does not only fix 
the couplings of the $Z'$ but also the interactions of the new Higgs states. In general, 
$U(1)$ extensions have a very interesting phenomenology: (i) they predict a $Z'$ boson which 
usually couples to SM fields; (ii) they could explain origin of $R$-parity and its 
spontaneous breaking \cite{FileviezPerez:2010ek,CamargoMolina:2012hv}; 
(iii) the absence 
of gauge anomalies predicts often right handed neutrinos and introduces therefore neutrino masses; 
(iv) many new dark matter candidates appear which not necessarily rely on the annihilation mechanisms known from 
the MSSM \cite{Basso:2012gz}; (v) the cross section of SUSY particles change compared to the MSSM
\cite{Krauss:2012ku}; (vi) 
$U(1)$ extensions might help to resurrect gauge mediated SUSY breaking (GMSB) \cite{Mummidi:2013hba,Krauss:2013jva}. \\
Even if the new scalars $\PH$ to break the gauge additional gauge symmetries are gauge singlets
under the SM gauge groups, the 
superpotential and the interactions with the MSSM doublets can be very different compared to the 
NMSSM because terms like $\PH^3, \PH \Hd\Hu$ are forbidden by the new gauge symmetry. However, 
$D$-term interactions between both sectors can even arise due to kinetic mixing even if the Higgs
and $\PH$ fields are not charged under the same gauge groups \cite{Holdom:1985ag}. Kinetic mixing will always be generated by
RGE running if the two $U(1)$s are not orthogonal \cite{Fonseca:2013jra}, but it has only a moderate effect on the 
Higgs masses and couplings \cite{O'Leary:2011yq}. 
The effects are more pronounced if there are direct $D$-term interactions 
like in $U(1)_R \times U(1)_{B-L}$ models \cite{Hirsch:2011hg,Hirsch:2012kv}. 
In this case the additional  $D$-terms as well as the mixing
with additional light scalars can give a large push to the Higgs mass as shown in Fig.~\ref{fig:U1R}. 

\subsubsection{$SU(N)$ extensions}
\begin{figure}[hbt]
\includegraphics[width=0.37\linewidth]{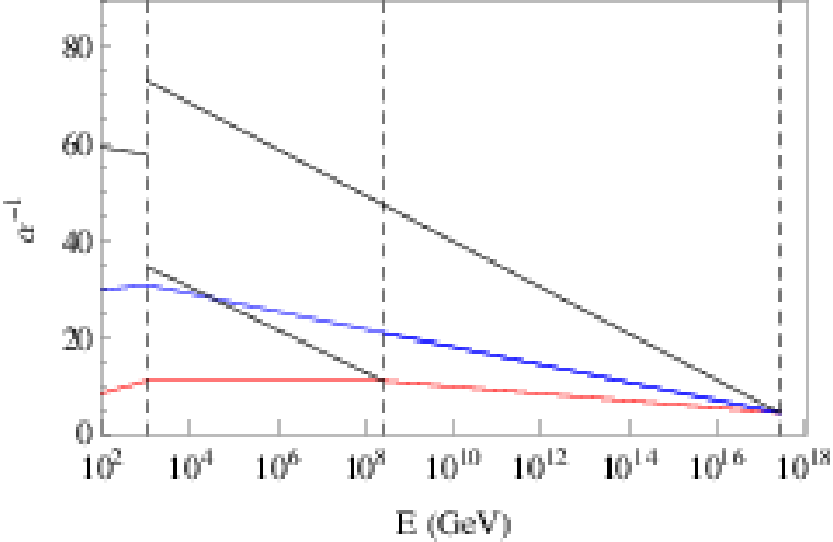} \hfill
\includegraphics[width=0.60\linewidth]{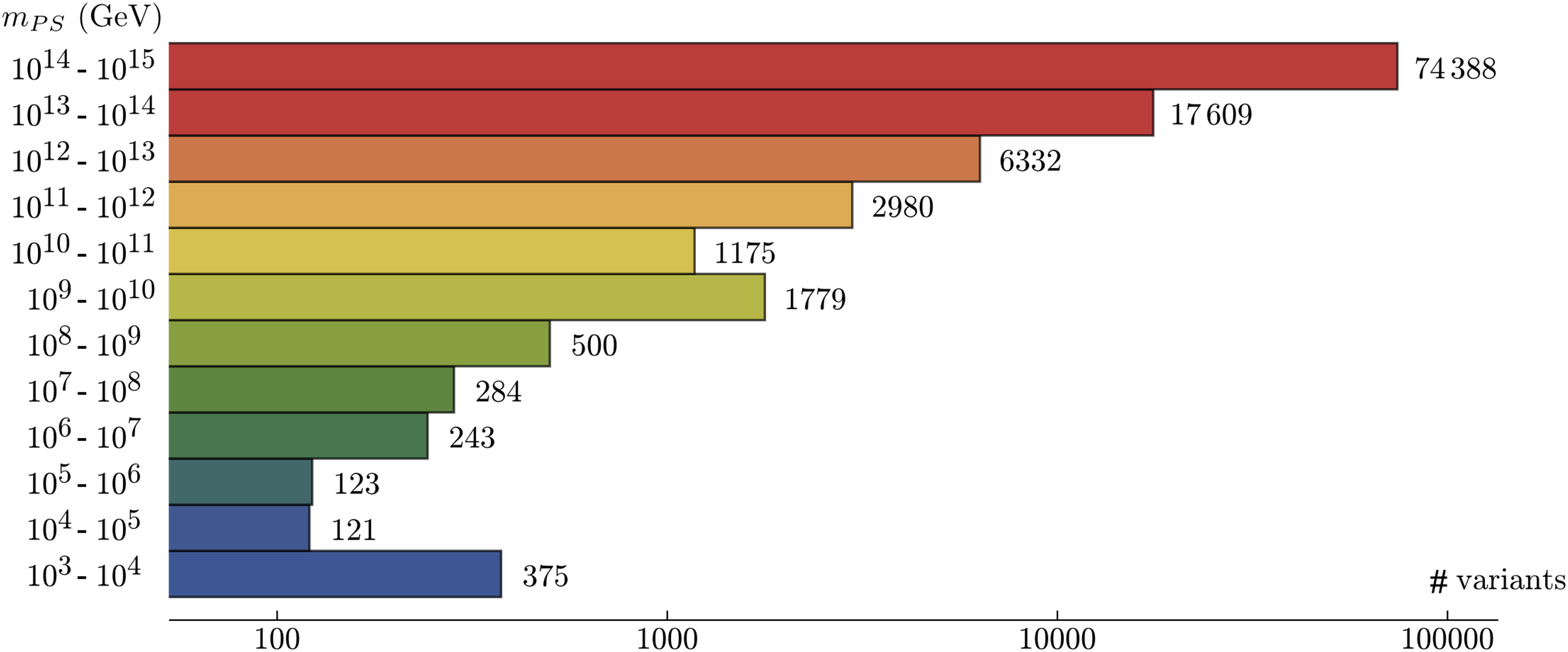} 
\caption{Left: Gauge coupling unification in $SO(10)$ models with 
an intermediate Pati-Salam scale. Plot taken Ref.~\cite{DeRomeri:2011ie}. 
Right: number of possible realization of such a model depending on the 
energy scale of the intermediate scale. Plot taken from \cite{Arbelaez:2013hr}.}
\label{fig:SUN}
\end{figure}

Additional $SU(N)$ are often motivated by $SO(10)$ GUTs. The GUT groups gets broken down to 
 to the SM gauge sector via the cascade
\begin{eqnarray*}
SO(10) &&\to SU(4)_{PS} \times SU(2)_L \times  SU(2)_R\\ 
&&\to SU(3)_c \times SU(2)_L \times  SU(2)_R \times U(1)_{B-L} \\
&&\to SU(3)_c \times SU(2)_L \times  U(1)_R \times U(1)_{B-L}  \to G_{SM}  
\end{eqnarray*}
Not always all intermediate steps are realized but it can be also be assumed that several steps happen 
at the same scale. This leads to three categories of models \cite{DeRomeri:2011ie}:
(i) Pati-Salam models, (ii) $SU(2)_L \times SU(2)_R$ models, (iii) $SU(2)_L \times U(1)_R$ models. 
Is has been shown that for each category 
many possible realizations exist which are consistent with gauge coupling unification, neutrino masses and a non-trivial CKM matrix
\cite{Arbelaez:2013hr}. See Fig.~\ref{fig:SUN}. All of these 
models have a very rich phenomenology because they predict  many new states together with a $Z'$ and $W'$s. In particular, new charged or even 
double charged Higgs bosons are a widely spread feature in these models. 

The double charged Higgs bosons in left-right models have been studied to some extent and mass limits of $M_{H^{++}} > 445~(409)$~GeV [CMS ~ (ATLAS)]
have been obtained \cite{CMS:2012ulp,ATLAS:2012hi}. Interestingly, there are also indirect constraints possible because $M_H^{++}$ 
can be correlated with $\delta\rho$ \cite{Bambhaniya:2014cia}.
For direct searches for double charged Higgs bosons multi-lepton channels are very promising \cite{Bambhaniya:2013wza}.

\section{Don'ts and Dos}
\label{sec:tools}
There is sometimes a huge difference in the manner how a BMSSM study is performed compared to the MSSM. Therefore, I want to comment 
on some aspects of the analyses and list public computer tools which should be considered to be used to bring BMSSM studies to a level 
comparable with MSSM standards. 
\begin{itemize}
 \item {\bf Tree-level Higgs masses in BMSSM models are not sufficient!} It is well known that the measured Higgs mass rules out 
 large areas of the parameter space of the (natural) MSSM. Thus, also the Higgs sectors of BMSSM models have to be confronted 
 with these limits. 
Thus, at least an one-loop calculation is mandatory to see if the Higgs mass is pushed into the correct direction.  If the one-loop 
mass turns out to be well below 120~GeV, it makes no sense to further study that parameter  point. 
One-loop calculations for a large range of BMSSM models can be either performed with \FeynArts/\FormCalc
 \cite{Hahn:1998yk,Hahn:2000kx,Nejad:2013ina}. Also the package \SARAH  \cite{Staub:2008uz} 
 together with either \SPheno \cite{Porod:2003um,Porod:2011nf} or \FlexibleSUSY \cite{Athron:2014yba} can be used 
 what provides a highly automatized calculation of loop masses.
 \item {\bf To get MC model files don't hack the MSSM one.} There are well established tools like
  \LanHEP \cite{Semenov:1998eb}, 
  \FeynRules \cite{Christensen:2008py,Alloul:2013bka}, or \SARAH to create model files for many Monte Carlo tools. 
  \item {\bf \HB/\HS \cite{Bechtle:2008jh,Bechtle:2013wla,Bechtle:2013xfa} should always be used 
  to check existing limits from Higgs searches and to give a quantative measure how good experimental data is reproduced.}. 
  These codes are generic enough to deal with highly extended Higgs sectors if the user provides the necessary input. 
  \item {\bf Check the vacuum stability.} It has been shown that the MSSM with light stops but a large mixing to explain the Higgs mass suffers 
  from an unstable, and often short-lived electroweak vacuum 
  \cite{Camargo-Molina:2013sta,Blinov:2013fta,Chowdhury:2013dka,Camargo-Molina:2014pwa,Chattopadhyay:2014gfa}. 
  To check the stability of the desired vacuum, the tool
 \Vevacious \cite{Camargo-Molina:2013qva} was created. 
  \item {\bf Don't forget about flavor physics.} Especially light, charged Higgs particles can be dangerous because they can significantly enhance observables 
  like $b \to s\gamma$.  The \FlavorKit interface \cite{Porod:2014xia} allows to calculate many flavor observables in BMSSM models 
  via the combination \FeynArts/\FormCalc -- \SARAH -- \SPheno. Alternatively, one can also use \FeynArts \& \FormCalc either stand-alone or coupled
 to {\tt Peng4BSM} \cite{Bednyakov:2013tca}.
\end{itemize}
One easy possibility for a precise study of BMSSM models is to use the {\tt SUSY} or {\tt BSM Toolbox} \cite{Staub:2011dp}. 
This is a collection of scripts 
which creates an environment consisting of \SARAH, \SPheno, \WHIZARD \cite{Kilian:2007gr,Moretti:2001zz}, 
\MG \cite{Alwall:2011uj,Alwall:2014hca} \HB/\HS, 
\CalcHep \cite{Pukhov:2004ca,Boos:1994xb}, \MicrOmegas 
\cite{Belanger:2001fz} and \SSP for the study of extended 
SUSY and non-SUSY models. Many of the models shown here are already delivered with \SARAH  and can be automatically implemented in all other tools
 via the {\tt Toolbox} scripts. 
In this context the \SPheno modules created for the new models provide a precise mass spectrum calculation based on two-loop RGE running and 
full one-loop corrections to all masses. An extensions for even a two-loop calculation in the Higgs sector is expected to appear soon
\cite{TwoLoopMasses}. \SPheno does also calculate two and three body decays for the  SUSY states present in the models 
and makes predictions for many flavor observables based on a full one-loop calculation. The scripts can be downloaded here
\begin{center}
{\tt http://sarah.hepforge.org/Toolbox.html}
\end{center}

\section{Conclusion}
I have briefly summarized the main aspects of SUSY models beyond the MSSM. One can see that there are many well motivated
 possibilities to go  beyond the MSSM. Each of the presented model has its peculiarities. While some models are already studied 
in great detail, others lack from a deep exploration. However, there are nowadays the tools available to perform precise studies 
in all models and to confront these models with experimental and theoretical constraints. 

\section*{Acknowledgments}
I thank the organizers of CHARGED'14 for the invitation and the hospitality during the stay. 
I'm supported by the BMBF PT DESY Verbundprojekt 
05H2013-THEORIE `Vergleich von LHC-Daten mit supersymmetrischen Modellen'.

\end{document}